\documentclass{interact}
\usepackage{lineno,hyperref}
\usepackage{listings}
\usepackage[dvipsnames]{xcolor}
\usepackage{lstbayes} 
\usepackage{afterpage}
\usepackage{hyperref}
\usepackage{amsmath}
\usepackage{float}
\usepackage{amssymb}
\usepackage{placeins}
\usepackage{amsmath}
\usepackage{hyperref} 
\usepackage{ctable} 
\usepackage{multirow} 

\modulolinenumbers[5]
\numberwithin{equation}{section}
\usepackage{pdflscape} 

\hypersetup{
    colorlinks=true,
    linkcolor=black,
    filecolor=magenta,      
    urlcolor=cyan,
}
\renewcommand{\vec}[1]{\mathbf{#1}}

\graphicspath{{./supplemental_online_materials/}}

\usepackage[natbibapa,nodoi]{apacite}
\begin{document}
\title{Confronting Quasi-Separation in Logistic Mixed Effects for Linguistic Data: A Bayesian Approach.\\ .\\ Draft of revisions as submitted to the Journal of Quantitative Linguistics. please cite published version from \url{https://doi.org/10.1080/09296174.2018.1499457}}

\author{
\name{Amelia E. Kimball\textsuperscript{a}, Kailen Shantz\textsuperscript{a}, Christopher Eager\textsuperscript{a} and Joseph Roy\textsuperscript{a}\thanks{CONTACT Amelia Kimball Email: akimbal2@illinois.edu}}
\affil{\textsuperscript{a}University of Illinois at Urbana Champaign} }

\maketitle
\begin{abstract} 

Mixed effects regression models are widely used by language researchers.  However, these regressions are implemented with an algorithm which may not converge on a solution. While convergence issues in linear mixed effects models can often be addressed with careful experiment design and model building, logistic mixed effects models introduce the possibility of separation or quasi-separation, which can cause problems for model estimation that result in convergence errors or in unreasonable model estimates. These problems cannot be solved by experiment or model design. In this paper, we discuss (quasi-)separation with the language researcher in mind, explaining what it is, how it causes problems for model estimation, and why it can be expected in linguistic datasets. Using real linguistic datasets, we then show how Bayesian models can be used to overcome convergence issues introduced by quasi-separation, whereas frequentist approaches fail. On the basis of these demonstrations, we advocate for the adoption of Bayesian models as a practical solution to dealing with convergence issues when modeling binary linguistic data.

\end{abstract}

\section{Introduction} \label{introduction}
Mixed effect models are commonly used in language science (e.g. Baayen, et al., 2017; Barr, et al., 2013; Gries, 2015; Matuschek et al., 2017). However, in much of the current debate about models for linguistic data, the differences between linear and logistic models is not discussed. In this paper, we highlight an issue which can impact logistic models but which is not present for linear models: separation and quasi-separation. These terms describe when one predictor completely or almost completely separates the binary response in observed data (see section \ref{separation} below). For example, in a discrimination task where native and non-native speakers are asked to judge whether auditory stimuli are the same or different, if all native speakers respond “different” and all non-native speakers respond “same” then the predictor of language background (native/non-native) will perfectly predict the outcome variable (same/different), resulting in separation. In this same task, if all native speakers respond “different”, but the non-native speakers sometimes respond “same” and sometimes respond “different”, we would have quasi-separation because the predictor of language background allows us to perfectly predict the outcome variable for one level (i.e. native speakers), but not for all levels of the predictor language background (because non-native speakers used both responses).  Separation and quasi-separation are common and expected in linguistic data (see section \ref{qsep_in_ling_data} below), and they render estimation of regression impossible or wholly unreliable using the frequentist methods which are widespread in linguistics \citep{levy2012probabilistic}. As a consequence, models may fail to converge and further yield unreasonable parameter estimates. 
\nocite{baayen2017cave}
\nocite{barr2013random}
\nocite{Gries:2015}
\nocite{matuschek2017balancing}

Our goal in this paper is to present Bayesian models as a means of dealing with quasi-separation in linguistic data. We first provide a general discussion  of convergence, after which we introduce Bayesian models and describe how these can provide a solution for modeling datasets containing quasi-separation, whereas frequentist models are likely to fail. 

Much of the literature on mixed effects models in language science has focused on whether to use 'Maximal Random Effects' (MRE). Barr and colleagues (Barr, et al., 2013; Barr, 2013), for example, recommend using MRE models as a way to minimize Type 1 error, while Baayen et al. (2017) and Matuschek et al. (2017), among others, argue against MRE. This debate, however, does not address the differences between linear and logistic models.  Rather than arguing for or against MRE, we discuss the problem of (quasi-)separation, which may be present in both maximal or non-maximal models, and present a Bayesian approach as a solution. 

\nocite{barr2013random,barr2013solo,baayen2017cave,matuschek2017balancing}

To demonstrate the utility of a Bayesian approach for overcoming quasi-separation, we present two linguistic datasets containing quasi-separation. In line with current advances in statistical methodology for linguistic data (\citealt{TQMP12-3-175}; Nicenboim \& Vashishth, 2016; Levy, 2012), we show how a Bayesian approach can achieve convergence where the frequentist approach fails, and further demonstrate that the failure of the frequentist models to converge is not likely due to low or zero variance, but rather due to the quasi-separation in the data. We conclude that Bayesian models provide a practical solution for handling quasi-separation in linguistic data, and argue that these therefore provide a better default strategy for modeling binary dependent variables.
\nocite{nicenboim2016statistical}
\nocite{levy2012probabilistic}

\subsection{Convergence}
The focus of our paper is the logistic mixed effects model as defined in (\ref{eqn:logmm})

\begin{equation}\label{eqn:logmm} 
\textrm{\textit{y} $\sim$ Bernoulli(\textit{p}), \qquad \qquad         log}\left(\frac{p}{1-p}\right) = \vec{X}\beta + \vec{Zu}
\end{equation}

The model represents the logistic function of \emph{p}, the proportion of our binary response, as a linear function of what are termed fixed effects, $\beta$, and random effects, $u$. The fixed effects are our predictors of interest while random effects are typically ancillary factors that the researcher is drawing repeated measures on, and for which the researcher wants to control the variance introduced by the repeated measurements (e.g. participants or items). Both $X$ and $Z$ are observed in a study, and  we are interested in estimating the parameters $\beta$ and $u$ from the data along with the covariance matrix introduced by the random effects, $\Sigma$.  We use the term \emph{parameters} to indicate terms in our model that are estimated from the data throughout this paper.  

In ordinary least squares regression for linear data, finding the estimate of the effect of a predictor involves solving an equation, and so arriving at an answer is straightforward.  Specifically, when $\boldsymbol{\beta}$ is estimated, that estimation process involves maximizing the likelihood for $\boldsymbol{\hat{\beta}}$ and yields $\vec{(X'X)^{-1}X'y}$ (in matrix notation).  Importantly, the estimated variance for each estimated predictor is also a closed calculation: $Var(\hat{\beta}) =\hat{\sigma}^2\vec{(X'X)^{-1}} $.  The only problem that arises to compute these formulae is when the data are collinear and thus the inverse of $\vec{(X'X)}$ (that is, $\vec{(X'X)^{-1}}$) does not exist.  

Moving from linear to logistic regression introduces the notion of convergence into the regression model. Through the 1970's, a unified approach to non-normal data was developed that resulted in what are termed generalized linear models (GLM) by McCullagh and Nelder (1983).\nocite{McCulloghetal1983} A GLM transforms the non-normal dependent variable via a link function, (as in equation  \ref{eqn:lgr1}) to be a linear combination of the predictors \citep{hardin2007generalized}.

\begin{equation}
\textrm{log}\left(\frac{p}{1-p}\right) = \vec{X}\beta
\label{eqn:lgr1}
\end{equation}

The estimation of a GLM is fundamentally different from OLS linear regression. Rather than being a closed equation with a single identifiable answer, GLM models are implemented with an iterative algorithm to solve for $\beta$,  and are said to \emph{converge} when the difference between two consecutive iterations of the algorithm is less than a pre-defined tolerance (see \citealp {monahan2011numerical} or \citealp {lange2010numerical} for a comprehensive technical overview). While there is a unique maximum likelihood estimate, no closed form solution exists, and so these algorithms can run into problems that result in a failure to converge. When we add random effects into a logistic regression model, achieving convergence is even more difficult.\footnote{Due to space constraints, we leave out a full technical discussion of what convergence means for these models (and Bayesian models).}  Convergence tests, essentially, are tests of reliability and reasonableness of the statistical results. These can vary across software implementations and versions of software packages. Crucially, convergence errors typically lead to a reduction in the random effects model by researchers on the assumption that the larger random effects model is incorrect, but convergence errors do not test model correctness.  The three main sources of convergence problems for logistic mixed effects models are model identifiability, low or zero variance, and separation. Each of these are discussed below.

\subsection{Model Identifiability}
Model identifiability  means that we can uniquely estimate our model parameters given a study design and the observed data. If a model is non-identifiable, no solution is possible, and the algorithm will fail to converge. 

There are two main conditions for identifiability in mixed effects models. First, there must be no collinearity in the random effects for at least one level of each random effect. That is, if regressions were run on the random effects structure for each subject individually, there needs to be at least one subject where the predictor matrix is full rank. Second, there must be more observations than parameters in the model. These formal requirements are a low bar, and most researchers will fulfill these conditions as a matter of course. In fact, identifiability can be assessed for linear mixed effects models prior to collecting any data (see Demidenko, 2013:117-120). For logistic mixed effects regression, however, identifiability is conditioned not only on the study design and the model design, but also on the observed $y$ (i.e. the distribution of the binary responses). Because of this, we can only assess the identifiability of a model once the data have been collected since if the responses show \emph{separation} (see section \ref{separation}), a model for even the most carefully designed study will not be identifiable, and will therefore not converge.
\nocite{Demidenko:2013}

\subsection{Low or zero variance}
A common cause of convergence errors in a frequentist mixed effects model is low or zero variance in some component of the random effects structure. Near-zero variance makes a random slope or intercept difficult to estimate without a very large set of data. Furthermore, in a frequentist approach this variance can only be reliably estimated in a converged model. This is a statistical analyst's catch-22: if the model does not converge, this may be due to low variance, but this cannot be assessed in a model that does not converge.  Bayesian models, in contrast, help to avoid this problem in two ways. First, simulation work has shown that convergence is up to 82 times more likely under a well-specified Bayesian model for logistic data compared to frequentist models \citep{chrisjoeLSAposter}. Second, even if a model does not converge, the Bayesian approach nonetheless makes it possible to assess the presence of zero or near-zero variance through examination of the posterior distribution.

\subsection{Separation \& Quasi-Separation in Logistic Models}   \label{separation}
In a logistic mixed effects model, whether a unique solution for $\beta$ exists is conditioned on  how the data are distributed \citep{ensoy2015separation}. For logistic models, three situations exist with respect to the observed distribution of the response variable and predictors: separation, quasi-separation and overlap (Ensoy et al., 2015; Allison et al., 2004)\nocite{allison2004convergence}. \emph{Separation} is when a predictor completely classifies the binary response in the data or when a covariate completely separates the data into a proportion of 1.0 or 0.0 for the response.  \emph{Quasi-separation} is when one level of a predictor or a covariate separates the data into a proportion of 1.0 or 0.0 for the response, though other levels do not show separation.  Finally, the ideal situation is termed \emph{overlap}, where the proportion of response is less than 1.0 and greater than 0.0 for every level of each predictor and the entire range of the covariates

In order to illustrate this, let's assume that there is a design where we want to estimate the effect of two conditions, A and B, on the accuracy of participants. For simplicity, we assume no repetition in either items or participants. Table \ref{tab:catSep} shows a  hypothetical experiment with 50 observations (25 per condition). The p(Condition) is the observed proportion of N in each condition.  Distributions reflecting separation, quasi-separation, and overlap are displayed. An example for a continuous predictor,  is shown in Table \ref{tab:covSep}. In both cases, under separation, the predictor displayed would perfectly predict the binary response.

[INSERT TABLE \ref{tab:catSep} ABOUT HERE]

[INSERT TABLE \ref{tab:covSep} ABOUT HERE]

 Examining data for (quasi-)separation can be relatively straight-forward for fixed effects structures: simply calculate the proportion of observations for each level of each fixed effect. For large studies with multiple predictors in the random effects structure, however, it can become very difficult to assess quasi-separation, as we will show with the second dataset we present below. Quasi-separation is problematic for frequentist models because it can make parameters difficult to estimate using maximum likelihood estimation. If, on top of that, there are random effects with low variance, quasi-separation can make convergence impossible, even with an overwhelming amount of data.

A common approach to dealing with failures to converge in the language sciences is to simplify the random effects structure until a model converges, but this technique has three major weaknesses:
\begin{enumerate}
\item{Unlike in typical model comparison, the model is being simplified not because the simpler model actually better captures the data being modeled, but because it is easier for the algorithm to estimate.}
\item{Running many models searching for convergence may allow for cherry picking or inflated error rates, especially when multiple choices of reducing the random effect structure provide convergence.}
\item{This technique does not allow a researcher to fully pre-register their statistical analysis.}
\end {enumerate}

Below, we will illustrate how a fully specified Bayesian model can achieve convergence in the face of quasi-separation, allowing researchers to avoid ad-hoc model simplification techniques when frequentist models fail.

\subsection{Quasi-separation and linguistic data}
\label{qsep_in_ling_data}
Before presenting a Bayesian approach, we briefly consider some types of linguistic data in which quasi-separation is likely to surface, in order to highlight the fact that quasi-separation is not restricted to the types of data we present in this paper, but rather should be expected across a range of linguistic datasets. Some examples of where quasi-separation could be expected in linguistic data are:

\begin{enumerate}
\item{In second language  research, where native populations and highly advanced learners often exhibit ceiling performance.}
\item{Research using grammaticality judgments or truth value judgments in which a particular condition elicits only one response type in a subset of participants or in which specific items may elicit a unanimous response across all participants.}
\item{Psycholinguistic experiments using a factorial design with a binary outcome measure, such as lexical decision accuracy, or probability of regression, fixation or skipping in eye-tracking research.}
\item{In sociolinguistic data where the likelihood of linguistic variables for some conditioning factors is low and the few tokens which do appear only occur with one variant.}
\item{In corpus data where linguistic forms may occur infrequently in some conditions of interest.}
\end{enumerate}

In short, quasi-separation can be expected in many types of linguistic datasets where the outcome variable is binary. The Bayesian approach we outline below may therefore be worth adopting for researchers in a variety of subfields across linguistics who may be confronted with issues of quasi-separation.

\section{Bayesian Approach to Logistic Mixed Effects}
Most statistical models reported in language science are \emph{frequentist} in nature. Frequentist models provide the probability of the observed data given the null hypothesis. If the probability, given by a \emph{p-value}, is below a predefined threshold, we reject the null hypothesis. The Bayesian approach, in contrast, provides the probability of a certain hypothesis given the observed data. For readers who are new to Bayesian models, we recommend \cite{kruschke2014doing}, \cite{nicenboim2016statistical} and \cite{TQMP12-3-175} for detailed introductions to Bayesian statistics.

\subsection{The problem of separation in frequentist models}
Frequentist approaches to mixed effects (e.g. \emph{lme4}) \citep{bates2014fitting} differ from Bayesian models in the way they estimate their results. In frequentist approaches, all possible values for the fixed effects $\beta$ and the random effects covariance matrix $\Sigma$ are considered equally likely \emph{a priori}, and the random effects coefficients $u$ are assumed to be multivariate normal distributed with mean zero and covariance $\Sigma$. For any given set of parameter values in a frequentist model, the probability of the observed data conditioned on the parameter values can be computed and is called the likelihood function. The goal of maximum likelihood estimation, which drives frequentist mixed effects models, is to find the set of parameter values which jointly maximize the value of this function. When (quasi-)separation occurs, maximum likelihood estimates can become extreme in their magnitudes, which can lead to unreasonably large estimates for $\beta$, $u$, and $\Sigma$. In practical terms, this means that (quasi-)separation will cause problems for the maximum likelihood estimates, making convergence less likely for frequentist models.

\subsection{Constraining a model with weakly informative priors}
Unlike frequentist models, Bayesian models do not assume that all values of $\beta$ and  $\Sigma$ are equally likely. Instead, Bayesian models incorporate constraints on the model parameters. For example, rather than assuming all values of  $\beta$ from negative infinity to infinity are equally likely, a model could begin with a starting assumption that values near zero are more likely. The analyst defines what these starting assumptions, or \emph{priors}, are. By defining priors, Bayesian models can decrease the likelihood of overestimating effect magnitudes when separation or quasi-separation occurs. Of course, the analyst must be careful to choose priors that will not bias their results. The type of constraints that we use here are called “weakly informative priors”, and are designed to focus on likely outcomes, while remaining conservative and avoiding undue bias \citep{gelman2008weakly}.

What the Bayesian approach offers us is a way to incorporate some common sense into the model: it is reasonable to assume that large effects are less likely than small effects \emph{a priori}, and that, on the log-odds scale, values outside of [-5, 5] are unlikely (but still possible). When there is enough data showing evidence of a large effect, we want the large effect to still be possible, but we do not want to consider an effect size of 6 equally likely to an effect size of 1 on the log-odds scale. This does not preclude the final estimates from residing outside the prior window -- it just requires strong evidence in the data for the appearance of large effects.

Mathematically, when applied to the fixed effects, these assumptions are expressed via a symmetric bell-shaped distribution with a peak at zero. We choose the Cauchy distribution, which has a similar shape to the normal distribution, but with much more probability in the tails, to the point where the distribution has no mean or variance, instead it has a location and scale \citep{gelman2008weakly}. By treating more extreme values as unlikely, a Bayesian model helps with locating effect estimates in the model of interest and prevents many convergence problems that originate in quasi-separation.

\section{Method and Data}

In what follows, we give two examples of mixed effects models with binary dependent variables. In both cases, previous research provides strong justification for a limited set of predictors and interactions, but convergence errors and quasi-separation create problems for the desired analyses when using a frequentist approach. Using these data, we will show how well-specified Bayesian models can overcome issues of quasi-separation to produce models that converge on the desired random effects structure.

\subsection{Dataset 1: Behavioral data from a psycholinguistic study}
The first dataset comes from the accuracy data reported in \cite{ShantzTanner2016}. Their study investigated the relative timing with which grammatical gender and phonological information are retrieved during lexical access. Specifically, they combined the use of ERPs with the dual-choice go/no-go paradigm in order to examine how task order impacts the relative timing with which grammatical gender and phonological information are retrieved during lexical access.
 
In their experiment, twenty native speakers of German were presented with 24 black-and-white images depicting high frequency, concrete German nouns. Nouns differed orthogonally in their grammatical gender and word-initial phone. Participants were presented with an image and asked to make a set of decisions based on the grammatical gender and the phonology of the depicted noun. One source of information (e.g. gender) was used to decide whether or not to respond (i.e. the go/no-go decision), and the other source of information (e.g. phonology) determined whether to respond with the left or right hand (i.e. the dual-choice decision). The mapping of information (i.e. gender or phonology) to decision (go/no-go or dual-choice) and response hand was fully counterbalanced within subjects; each possible configuration occurred in a separate experimental block. Within a block, items were presented four times each, yielding a total of 768 trials per participant, or 15,360 data points. To manipulate task order, the order of blocks was counterbalanced across four lists: two in which go/no-go decisions were based on phonology for the first half of the experiment and gender for the second half, and two in which go/no-go decisions were determined by gender for the first half of the experiment and phonology for the second half. In addition to task order, condition (i.e. hand = phonology or hand = gender), trial type (i.e. go or no-go) and the interactions between these variables were the predictors of primary interest. Trial type and condition were included in the accuracy model, as these were the two variables on the basis of which relative time course information was determined in the electrophysiological data. It was thus important to determine whether task order would impact accuracy as a function of these predictors.

The logistic mixed effects regression model fit to these data also included a number of control predictors. Trial was included in the model, given that performance in psycholinguistic experiments can change over the course of a session due to learning (e.g. \citealp*{fine2013rapid}) or fatigue. Grammatical gender was included as a control predictor, as prior research has shown effects of grammatical gender on lexical access \citep*{akhutina1999processing, bates1996gender, Opitz2016gender}. Because phone frequency has been found to influence errors in speech production \citep{levitt1985roles}, and because participants were required to retrieve information about each item's initial phone, the initial sound was further included as a control. Word frequency, moreover, has known effects in lexical access \citep*{jescheniak1994word, strijkers2009tracking, strijkers2011conscious}, and was therefore included in the model. Finally, the number of syllables in a word has also been shown to influence production in picture naming paradigms \citep{alario2004predictors}; syllable count was thus included as a control predictor. Because the task order manipulation split conditions across different halves the the experiment for each group, any interaction between condition and task order was potentially confounded by trial effects. The authors thus also included the three-way interaction of task order x trial x condition and its subordinate interactions to statistically control for this potential confound. Table \ref{kailenpredictors} summarizes the predictors included in the full model for \citet{ShantzTanner2016}. 

[INSERT TABLE \ref{kailenpredictors} ABOUT HERE]

\subsubsection{Potential for quasi-separation in the data}
Note that even before running the model, it is reasonable to expect quasi-separation in this type  of data. Indeed for speech to function efficiently, we should expect that lexical retrieval processes will generally proceed without error. While the production system is sensitive to task demands, it should also be robust to these such that retrieval errors do not drastically increase if task demands are increased. Thus, we should expect performance in tasks such as these to be at or near ceiling, and for any experimental effects to be relatively small.

Accordingly, examining the mean performance of each group (Table \ref{KailenAccuracySummary}) in each condition shows that accuracy was very near ceiling.  If we further examine these data  on a per-subject basis, we find that quasi-separation is present in these data for some, but not all participants. Table \ref{KailenAccuracySubset} illustrates this with data from four participants who show varying degrees of quasi-separation. A full summary for all participants can be found in the supplementary material.\footnote{all items referenced as "supplementary materials," including .rda files containing full code and summarized data, may be found at \url{https://www.doi.org/10.17605/OSF.IO/ZHUJF}}

[INSERT TABLE \ref{KailenAccuracySummary} ABOUT HERE]

[INSERT TABLE \ref{KailenAccuracySubset} ABOUT HERE]

\subsection {Dataset 2: A Perception Study}

The second study we examine is data from Kimball \& Cole (2018) . This study investigates the effects of phonological features and phonetic detail on the reported perception of stressed syllables in sentences of English. Recent experiments on prosody perception show that listeners use both signal-based cues (e.g. duration) and top-down cues (e.g. information structure) when reporting the perceived prominence of words in English (~\citealp{Bishop:2012}; \citeauthor*{ColeMoHJ:2010}, 2010). However, it is not known whether a similar array of factors influence perception of syllable stress. Kimball \& Cole (2018) report on a stress perception study that tests the interaction of acoustic cues with top-down cues related to metrical context and lexical stress location. 

\nocite{KimballColeBuckeye}

For the experiment, self-reported native speakers of American English were recruited online using Amazon Mechanical Turk. Kimball and Cole use a  stress reporting task in which participants are presented with an audio file along with a syllable-by-syllable transcription of that file, displayed in a browser window using Qualtrics survey software. Participants listened to recordings of sentences and marked the stressed syllables in the transcript.  In total, 94 subjects marked 403 syllables each for a total of 37,882 data points. 

The dependent variable in the logistic mixed effects regression model fit to these data was a binary variable denoting whether a syllable was marked as stressed or not. Fundamental frequency (or \textit{f0}, the acoustic measure that is perceived as pitch), intensity, and duration were included in the model as signal-based predictors, based on prior work suggesting that these are acoustic markers of prosodic prominence (\citealp{Bishop:2012}; \citealp{Breenetal:2010}; \citeauthor*{ColeMoHJ:2010}, 2010). Non-signal-based predictors included stress location in citation form (i.e. primary stress as it would be marked in the dictionary), which was chosen based on previous work showing that this is a reliable predictor of reported syllable stress \citep{KimballCole:2014}, as well as function/content word status, which was included as a predictor given that function words are less likely to attract stress \citep{Selkirk:1995}.

Metrical context was also included because prior work on stress shift (\citealp{Grabe:1995}; \citealp{Vogel:1995}) indicates that the metrical structure of English disprefers two adjacent stresses. This was incorporated in the model by including a predictor indicating whether the previous syllable was marked as prominent. Finally, the work presented here was originally run as two separate experiments with slightly different instructions but identical procedures; results of the two experiments are pooled in this model, and so experiment instructions is added as a predictor. Based on previous research, variability between subjects and items is expected on all of the predictors except experimental instruction (\citeauthor*{ColeMoBaek:2010}, 2010; \citealp{KimballCole:2014}), and so there is strong motivation to include these as random effects. Table \ref{ameliapredictors} summarizes the predictors included in the model reported by Kimball and Cole (2018).

[INSERT TABLE \ref{ameliapredictors} ABOUT HERE]

\subsubsection{Quasi-separation in the data}
Given previous results which indicate variation in reported stress location, there is the possibility that individuals use different cues to determine stress, or weight cues differently, which would lead to an expectation of separation for at least some participants in some conditions. \emph{A priori} there is no reason to expect--or to rule out--(quasi-) separation. 

A quick inspection of proportions of observations at each level in the fixed effects shows no obvious separation. Quasi-separation is, however, present in the random effects structure. This is not easy to see at a glance from descriptive measures, but can be shown using a classification tree (provided in the online supplementary material). The fact that a classification tree was necessary to identify the quasi-separation in this dataset illustrates how quasi-separation is not always easy to find in large, complex models. For a researcher who is not familiar with quasi-separation, or with data visualization techniques such as classification trees, this problem may present as intractable convergence errors with no obvious source. This highlights the fact that we cannot (and perhaps should not) assume an absence of quasi-separation in our data.

\subsection{Statistical Model Procedure}

In this section we provide a detailed description of the statistical procedure we followed and our reasoning behind the procedural choices we made.  For more detail on how to perform these procedures, see \citep{kruschke2014doing,nicenboim2016statistical}.  To perform or inspect our model, consult the shinystan objects, which contain the full posterior sample along with the Stan model code, in the supplementary online materials.

\subsubsection{Scaling}
In order to choose appropriate priors for the model, we need to first ensure that the regression predictors are themselves all on a similar scale. We thus used the \emph{standardize} function in the \emph{standardize} package \citep{eager2017standardize} to place all predictors on a scale of 0.5 (i.e. covariates have standard deviation 0.5, unordered factors have sum contrasts with deviations of +/- 0.5, and ordered factors have orthogonal polynomial contrasts with column standard deviations of 0.5). Choosing a scale of 0.5 allows us to easily extend the prior distributions discussed in Gelman et al. (\citeyear{gelman2008weakly}). In this way, interaction terms automatically have smaller scales than main effects based on the order of the interaction (Gelman et al., 2014).  \nocite{gelman2014bayesian} This requires a stronger signal to establish interaction effects in our analysis and reduces the chances of seeing an erroneous large estimate for an interaction term that is not supported by the data, but a by-product of an unconstrained model. 

\subsubsection{Selection and specification of priors}
Given the scaling above, we chose a scale of 4 for the Cauchy prior distribution on the fixed effects.  This places approximately 75\% of the probability density between -10 and +10.  Because the scale of the predictors is 0.5, this means that, before observing the data, we are assuming there is a 75\% probability that the effect in log-odds of a 1-SD increase in a covariate is between -5 and +5, and similarly there is a 75\% probability that the difference in log-odds between two groups in a binary unordered factor is between -5 and +5 (reflecting the reasonable assumption discussed above).  For interactions, our prior assumption is automatically adjusted such that we assume that interactions likely have smaller effect sizes than main effects (the actual scale of the prior is the same, but the reduced scale of the interaction terms cause large values for $\beta$ to actually have a smaller interpretation; e.g. for a second order interaction of unordered factors, the value of $\beta$ is always multiplied by +/- 0.25 or 0, etc.).  

For the fixed effects intercept term, we use a Cauchy distribution with location zero and scale 2.5, placing 75\% of the prior probability between -6 and +6 (i.e. the most likely case is that 0’s and 1’s are equally likely, and there is a 75\% chance that the corrected mean proportion of 1’s is between 0.0025 and 0.9975).  In doing this, we place  constraints on the effect sizes, while still allowing substantial prior probability for large effect sizes (25\% in each case) when the data warrant such effects.

For the standard deviations in the random effects, the effect of the predictor scales is the same, and the prior distributions are conceptually similar except that negative values are not possible. We therefore use a half Cauchy distribution (i.e. just the positive side of the bell-shaped curve centered at zero) with a scale of 2 for both random intercepts and random slopes. The median of a half Cauchy distribution is the scale parameter itself, and 75\% of the probability density is between 0 and 2.5 times the scale \citep{gelman2006prior}. In this way we are making the very weak assumptions that the standard deviation in the random intercepts is most likely less than 5 (75\% probability), and the standard deviation in the random slopes for an increase of one standard deviation in a covariate is most likely less than 2.5, and so on.)

For the correlations in the random effects, rather than considering all possible correlation matrices equally likely, we place a \emph{LKJ} prior on the correlation matrix with a shape parameter of 2 (\citeauthor{lewandowski2009generating}, \citeyear{lewandowski2009generating}). This prior has a peak at zero correlation (for the same reason that the fixed effects prior has a peak at zero effect), is symmetric (positive and negative correlations are equally likely), and favors smaller correlations over larger correlations.  It is most helpful to visualize this prior as a $\beta$(2,2) distribution on the range (-1, 1) rather than (0, 1).  When there are only two random effects (i.e. random intercept and one random slope), the correlation matrix only has one free parameter, and \emph{LKJ}(2) is the same as $\beta$(2,2) on (-1,1).  When there are more than two random effects, the requirement that correlation matrices be positive definite adds additional constraints to the possible values that the free parameters take, but conceptually the prior is the same, with 0 in the $\beta$(2,2) on (-1,1) distribution representing zero correlation (i.e. the identity matrix), and higher values representing greater or lower correlation in general. 

The random effects covariance matrix $\Sigma$ is deterministically related to the standard deviations (which we will denote with $\sigma$) and correlation matrix (which we will denote with $\Omega$): $\Sigma$ = diag($\sigma$)*$\Omega$*diag($\sigma$).  For the random effects coefficients  (i.e. the effects for the individual subjects and items) the same multivariate normal distribution assumption used in frequentist approaches is used. Letting $y$ be the observed response (0’s and 1’s), $X$ the fixed feature matrix (excluding the intercept) standardized to scale 0.5, $Z$ the random feature matrix (including intercepts), $\beta_0$ the fixed intercept, $\beta$ the fixed effects coefficients, $u$ the random effects coefficients, $\Sigma$ the random effects covariance matrix, $\sigma$ the random effect standard deviations, and $\Omega$ the random effects correlation matrix, our model constraints (i.e. prior assumptions) can be summarized as:

\begin{itemize}
\item $\beta_0 \sim Cauchy(0,2.5)$
\item $\beta \sim Cauchy(0,4)$
\item $\sigma \sim HalfCauchy(0,2)$
\item $\Omega \sim LKJ(2)$
\item $\Sigma = diag(\sigma)*\Omega*diag(\sigma)$
\item $u \sim MultivariateNormal(0,\Sigma)$
\item $y \sim Bernoulli(logit^{-1}(\beta_0 + X\beta + Zu))$
\end{itemize}
In order to incorporate these constraints, Bayes’ rule is used to arrive at the full probability distribution of the parameter values given the observed data, called the posterior distribution of the parameters. That is, we have \emph{Posterior} $\propto$ \emph{Prior * Likelihood}, where $\propto$ means \emph{is proportional to}. Actually solving the equation to arrive at a closed form for the posterior probability distribution is often impossible (as is the case with the regressions in this paper) or otherwise analytically intractable, and so the distribution is estimated algorithmically using a Markov Chain Monte Carlo (MCMC) technique.  The implementation of MCMC for logistic mixed effects regression is the same as the implementation for linear mixed effects regression (they simply differ in the prior on $y$).  For a thorough introduction to MCMC, we refer the reader to \citep{geyer2011introduction}  In this paper, we use the No U-Turn Sampler (NUTS), a specific type of Hamiltonian Monte Carlo (HMC)\citep{neal2011mcmc}, which is itself a special type of MCMC. We implement this in RSTAN, the R interface to the Bayesian programming language, Stan (Carpenter et al., \citeyear{carpenter2016stan}).

\subsubsection{Evaluating the Model Fit}
There are important differences in interpretation when using a Bayesian model versus a frequentist model.  Most notably, a frequentist model gives us specific values for $\beta$ which it determines is most likely, while the Bayesian model shows us a density plot of probable values of $\beta$, called the posterior distribution.  In the supplementary materials, we report the estimates for the posterior mean, standard deviation, median, and a 95\% credible interval (also called an uncertainty interval)\footnote{This approach requires the analyst to focus on each estimated parameter rather than the traditional approach in some fields for omnibus tests followed by post-hoc tests. Due to space requirements, we do not fully address this, but only choose to say that there are Bayesian forms of this approach which could be implemented, but are themselves controversial (i.e. Bayes Factors). In any case, the approach outlined here moves the researcher away from a \emph{p-value} only statistical analysis to examining more of the statistical model relevant for a research question.}.

For the regression models in this paper, we used the recommended defaults for the number of chains (4), the number of iterations per chain (2000, of which 1000 are warmup), and the target acceptance rate (\emph{adapt\_delta}, in the Stan language, set at 0.8).  To assess convergence, we ensured that there were no divergent transitions post-warmup (which would indicate that the target acceptance rate needs to be increased), that the Gelman-Rubin  statistics was under 1.1 for all parameters, that trace plots showed good mixing for the chains, that posterior density plots showed no multimodality, and that random draws from the posterior predictive distribution approximated the sample distribution. (\citealp{gelman1992inference}, \citealp{gelman2011inference}).

These convergence checks can be easily performed using the \emph{shinystan} package \citep{shinystan}. Because all $\hat{R}$ statistics were under 1.1, we omit them from the tables presented in the supplementary materials, but do include the effective sample size for each parameter in addition to the statistics described earlier (mean, SD, median (50\%), 95\% CI). 

\section{Results}
\subsection{Behavioral task} 

Results from Shantz and Tanner (2017) found a robust effect of task order (i.e. whether the go/no-go decision was determined by gender or by phonology first); the group of participants who used phonology to make the go/no-go decision in the first half of the experiment were significantly faster at responding, and showed electrophysiological evidence for earlier retrieval of grammatical gender over phonology, consistent with prior research \citep{vanTurennout:1998}. In contrast, the gender = go/no-go first group showed no evidence for earlier retrieval of gender. While the accuracy data are numerically consistent with this trend of group differences, showing higher accuracy for the gender = go/no-go first group, the mixed effects model fit to these data found no reliable main effect of task order. Importantly, however, this model did not converge with the random effects structure that the authors originally attempted to fit. In fact, the model reported in Shantz and Tanner (2017) was simplified down to only a random intercept by subject, and even then did not converge.

In contrast to the models fit using a frequentist approach, the Bayesian model we fit to these data converged on the desired random effects structure without the need for any form of model simplification. Examining the variance components, there is good evidence for non-zero variance in all of the random effects. Thus, the failure of the standard model fit with \emph{lme4} to converge is not likely attributable to a lack of variance by items or by subjects. Figure 1 shows the variance estimates of the random intercepts for subjects. Plots of the variance estimates for all other random effects can be found in the supplementary materials. 

[INSERT FIGURE 1 ABOUT HERE]

Given that quasi-separation is known to be present in the data, and that non-zero variance is present for all random effects, quasi-separation is thus implicated as the likely culprit for the failures of the frequentist models to converge. The fact that the Bayesian model converged on the desired random effects structure demonstrates the power of a Bayesian approach to solve issues due to quasi-separation while allowing researchers to fit their optimal random effects structure. The results of the Bayesian model for this dataset are shown in the supplementary materials. 

Importantly, the model does not find strong evidence for an effect of task order on accuracy, as the credible intervals for the effect of task order and the interactions of task order with trial type and condition all contain zero. In other words, the results of the Bayesian model indicate that task order did not impact accuracy. While this is ultimately in line with the modeling results reported in Shantz and Tanner (2017), the fact that the Bayesian model converged allows us to  confidently conclude that task order has no effect on accuracy, whereas Shantz and Tanner (2017) refrained from drawing any such conclusion due to the fact that their estimates were based on a non-converged and thus unreliable model. 

\subsection{Perception study} 

As was the case for Dataset 1, the frequentist model did not converge with the random effects structure that we built based on the literature, even though all available optimizers were attempted\footnote{The model failed to converge with a max gradient of 0.006. This is close enough to a 0.002 threshold that allowing the algorithm to estimate for, say, ten times longer might have led to convergence. However, the model specified here took 1 day to run. From a practical standpoint, we believe that advocating for models which would take ten days to run would be unreasonable when there is a solution in the form of Bayesian approaches. We also tried many different optimizers, none of which achieved convergence.}. Convergence was not achieved until the model was simplified to include only a random intercept by subject and no other random slopes or intercepts. In other words, the only frequentist model that converged did not account for random variance that should be expected based on prior research (e.g.~\citeauthor*{ColeMoBaek:2010}, 2010; ~\citealp{KimballCole:2014}). The Bayesian model, on the other hand, converged on the full randoms effect structure (for detailed results, see table provided in supplementary materials).
 
When the variance components from the Bayesian model are inspected, we find evidence for non-zero variance in all random effects. Thus, the lack of convergence with \emph{lme4} is most likely not due to low or zero variance by items or by subjects. To illustrate, Figure 2 shows the variance estimates of the random slope for primary stress by subject. Plots of the variance estimates for all other random effects can be found in the supplementary materials. The results of the Bayesian model for this dataset are shown in the supplementary materials. 
 
[INSERT FIGURE 2 ABOUT HERE]

This is exactly the type of vexing situation we would like to draw attention to: this model was carefully built, the experiment was designed to provide enough data to ensure adequate power, there was no reason \emph{a priori} to assume separation would be present, complete separation was not observed in any of the fixed effects, and still the frequentist models failed to converge (despite the fact that a post-hoc analysis shows non-zero variance in the chosen random effects). Given that the model is identifiable and has non-zero variance, quasi-separation is a likely culprit in the failures of the frequentist models to converge, and indeed it is present, as shown in the classication tree in the supplementary materials. 

\section{Discussion}
Most, if not all, researchers who use mixed effects models encounter convergence errors. When these errors occur, they must be addressed, as the results of a model that does not converge are invalid. The literature on linear mixed effects models provides some guidance on how to address convergence errors, however it tends to ignore critical differences between linear and logistic models. In this paper, we have highlighted one such difference, (quasi-)separation, and shown how Bayesian models can be used to overcome the issues introduced by (quasi-)separation.

Our findings show that a failure to converge should not be taken as a reliable diagnostic of poor model specification or low or zero variance in the random effects (contra Bates et al., 2015: p. 3).
\nocite{bates2015parsimonious}
Rather, we have argued that the quasi-separation present in both datasets is likely responsible for the failures of the frequentist models to converge. These results demonstrate the importance of assessing the presence of (quasi-)separation as a potential source of convergence errors in logistic mixed effects models. Although identifying quasi-separation can be straightforward, as with dataset 1, we have also shown with the second dataset that it can be difficult to identify quasi-separation. Given this, and the fact that quasi-separation is likely to be prevalent in linguistic datasets, we believe a Bayesian approach offers the best default strategy for modeling linguistic data with binary dependent variables. We do acknowledge that a switch to Bayesian models requires an investment of time and effort, and we do not intend to insist that all  researchers must make this switch. However, we believe that the long-term benefits outweigh the upfront investment in learning how to fit Bayesian models. 

In addition to retaining the advantages that mixed effects models have for modeling linguistic data over traditional approaches, a Bayesian approach offers a number of additional advantages. First, the increased likelihood of convergence with a well-specified Bayesian model means that researchers are more likely to be able to fit a desired model structure without the need to resort to ad-hoc model simplification techniques in the face of non-convergence. This would help to avoid issues that arise from model simplification, such as those we discussed in Section 1.4; it would further make it more likely that a researcher can implement a pre-registered analysis plan, which would facilitate increased transparency in research practices as the move toward open science continues. In addition, though some Bayesian models may take longer to run than their equivalent frequentist models, the fact that they are more likely to converge can save valuable researcher time that would otherwise be spent refitting models in attempts to achieve convergence. Finally, if a Bayesian model does fail to converge, the model output allows researchers to assess whether that failure to converge may be due to low or zero variance, whereas this is not possible with frequentist models. Thus, researchers can make principled, informed decisions about whether to exclude a random effect on the basis of the data, rather than simply assuming low or zero variance.

\section{Conclusion}
Mixed effects models have gained favor among language researchers for their numerous advantages over traditional methods of data analysis. These advantages, however, are often undercut by the problems introduced by convergence errors, which are particularly likely when modeling binary dependent variables. We have shown how quasi-separation can result in convergence failures when using frequentist logistic mixed effects regression, and how Bayesian models can be used to overcome these convergence issues. This finding highlights the importance of assessing the presence of (quasi-)separation when convergence errors are encountered, as a failure to converge cannot be taken as evidence for low or zero variance in a random effects parameter. In light of these findings, we argue that a Bayesian approach is a better default strategy for modeling binary dependent variables.

\section {Acknowledgements}
This work was supported by NSF grants BCS-1251343 to Jennifer Cole and Jose I. Hualde, BCS-1349110 and BCS-1431324 to Darren Tanner, and by a grant from the Illinois Campus Research Board (RB14158) to Darren Tanner. AK received financial support from an Illinois Distinguished Fellowship from the University of Illinois. KS received financial support from a Doctoral Fellowship from the Social Sciences and Humanities Research Council of Canada, and from an Illinois Distinguished Fellowship from the University of Illinois. This research was supported by equipment funded from the Office of the Vice-Chancellor of Research at the University of Illinois at Urbana-Champaign to JR. Thanks to Darren Tanner and Jennifer Cole for making their data available. 

\bibliography{JQL_Bayesian}

\afterpage{%
\begin{table}[H]
\caption{Illustration of Binary Response Distribution for a Factor}
\label{tab:catSep}
\begin{tabular}{|l|c|c|c|}
\hline
Type & p(Condition A) & p(Condition B) & N\\
\hline 
Separation & 1.0 & 0.0 & 50 \\
\hline
Quasi-Separation & 1.0 & .75 & 50 \\ 
\hline
Overlap & .80 & .65 & 50\\
\hline
\end{tabular}
\end{table}
\clearpage
}

\afterpage{%
\begin{table}[H]
\caption{Illustration of Binary Response Distribution for a Covariate}
\label{tab:covSep}
\begin{tabular}{|l|c|c|c|}
\hline
Type & p(Time $>$ 50ms) & p(Time $\leq$ 50ms) & N\\
\hline 
Separation & 1.0 & 0.0 & 50 \\
\hline
Quasi-Separation & 1.0 & .75 & 50 \\ 
\hline
Overlap & .80 & .65 & 50\\
\hline
\end{tabular}
\end{table}
\clearpage
}

\afterpage{%
\begin{table}
\caption{Summary of Predictors for Dataset 1}
\label{kailenpredictors}
\begin{center}
\footnotesize
\begin{tabular}{llll}
\multicolumn{4}{c}{Categorical Predictors}\\
\hline
Predictor & Levels & Variable Type\\
\hline
Task Order & 1.Phonology = Go/No-Go First & Dichotomous\\
&2.Gender = Go/No-Go First\\
Trial Type & 1.Go &Dichotomous\\
&2. No-Go&\\
Condition&1. Hand= Phonology & Dichotomous\\
&2. Hand =Gender\\
Grammatical Gender & 1. Masculine & Dichotomous\\
&2. Neuter&\\
Initial Sound & 1. /k/ & Dichotomous\\
&2./b/&\\
Syllable Count & 1.One & Ordinal\\
&2. Two&\\
&3. Three&\\
\\
\multicolumn{4}{c}{Continuous Predictors}\\
\hline
 &Range &Mean & s.d.\\
\cmidrule{2-4}\\
Trial & 1-768 & 384.500 & 221.847\\   
Log Frequency& 1-3.916&1.458&0.644\\
\\

\multicolumn{4}{c}{Interactions}\\
\hline
\multicolumn{4}{l}{Task Order x Condition} \\
\multicolumn{4}{l}{Trail Type x Condition}\\
\multicolumn{4}{l}{Task Order x Trial Type x Condition}\\
\multicolumn{4}{l}{Task Order x Trial x Condition}\\

\\
\multicolumn{4}{c}{Random Effects Structure}\\
\hline
Intercept Term & Slope Term\\
\hline
Participant & Condition\\
&Trial Type\\
&Condition x Trial Type\\
Item & Condition\\
& Task Order\\
& Trial Type\\
& Condition x Task Order\\
&Condition x Trial Type\\
&Task Order x Trial Type\\
&Task Order x Trial Type x Condition\\
\hline
\end{tabular}
\end{center}
\end{table}
\clearpage
}

\afterpage{%
\begin{table}
\caption{Mean Accuracy (\%) and Standard Deviation by Group, Condition and Trial Type}
\label{KailenAccuracySummary}
\begin{tabular}{lrrrr}
\hline
&
\multicolumn{2}{c}{Hand = Phonology}&
\multicolumn{2}{c}{Hand = Gender}\\
\cmidrule{2-5}
& Go & No-Go & Go & No Go\\
\cmidrule{2-5}
Phonology = Go/No-Go First & 97.7 (1.1) & 99.4 (0.8) & 97.1 (2.6) & 99.2 (0.9) \\
Gender = Go/No-Go First& 98.3 (2.4) & 98.9 (1.8) & 98.2 (1.8) & 99.7 (0.5)\\
\hline
\end{tabular}
\end{table}
\clearpage
}

\afterpage{%
\begin{table}
\caption{Accuracy (\%) by Condition and Trial Type for a Subset of Participants to Illustrate Quasi-Separation in the Data}
\label{KailenAccuracySubset}
\begin{tabular}{lrrrr}
\hline
&
\multicolumn{2}{c}{Hand = Phonology}&
\multicolumn{2}{c}{Hand = Gender}\\
\cmidrule{2-5}
& Go & No-Go & Go & No Go\\
\cmidrule{2-5}
Participant 201 & 99.5 & 100 & 97.4 & 100\\
Participant 202 & 100 & 100 & 99.5 & 100\\
Participant 203 & 96.9 & 99.5 & 97.4 & 99.5\\
Participant 205 & 92.1 & 96.9 & 97.3 & 100\\
\hline
\end{tabular}
\end{table}
\clearpage
}

\afterpage{%
\begin{table}
\caption{Summary of Predictors for Dataset 2\label{ameliapredictors}}
\begin{center}
\small
\begin{tabular}{llll}
\multicolumn{4}{c}{Categorical Predictors}\\
\hline
Predictor & Levels & Variable Type\\
\hline
Stress & 1.primary stress  & Dichotomous\\
&2.no primary stress &\\
Function word & 1.Yes &Dichotomous\\
&2. No&\\
Word to the left marked&1. Yes & Dichotomous\\
&2. No&\\
Word to the right marked & 1. Yes & Dichotomous\\
&2. No&\\
Experiment instructions & 1. 'mark the beat' & Dichotomous\\
&2.'mark the syllable'&\\
\\

\multicolumn{4}{c}{Continuous Predictors}\\
\hline
&Range &Mean & s.d.\\
\cmidrule(lr){2-4}\\
F0(Hz) & 77.41-372.47 & 177.14 & 37.840\\
Intensity (dB) &  47.44-76.20  & 64.22  & 4.685 \\
Duration (ms) &  40.8-848.0 & 277.4 & 113.994  \\
\\

\multicolumn{4}{c}{Interactions}\\
\hline
\multicolumn{4}{l}{F0 x Intensity} \\
\multicolumn{4}{l}{F0 x Duration}\\
\multicolumn{4}{l}{Duration x Intensity}\\
\multicolumn{4}{l}{F0 x Duration x Intensity}\\
\\

\\
\multicolumn{4}{c}{Random Effects Structure}\\
\hline
Intercept Term & Slope Term\\
\hline
Subject  & Stress\\
&Function word\\
&F0\\
&Intensity\\
&F0xIntensity\\
Item & Experiment\\
& Previous word marked\\

\end{tabular}
\end{center}
\end{table}
\clearpage
}

  \begin{figure}[H]
       \includegraphics[width=12cm]
       {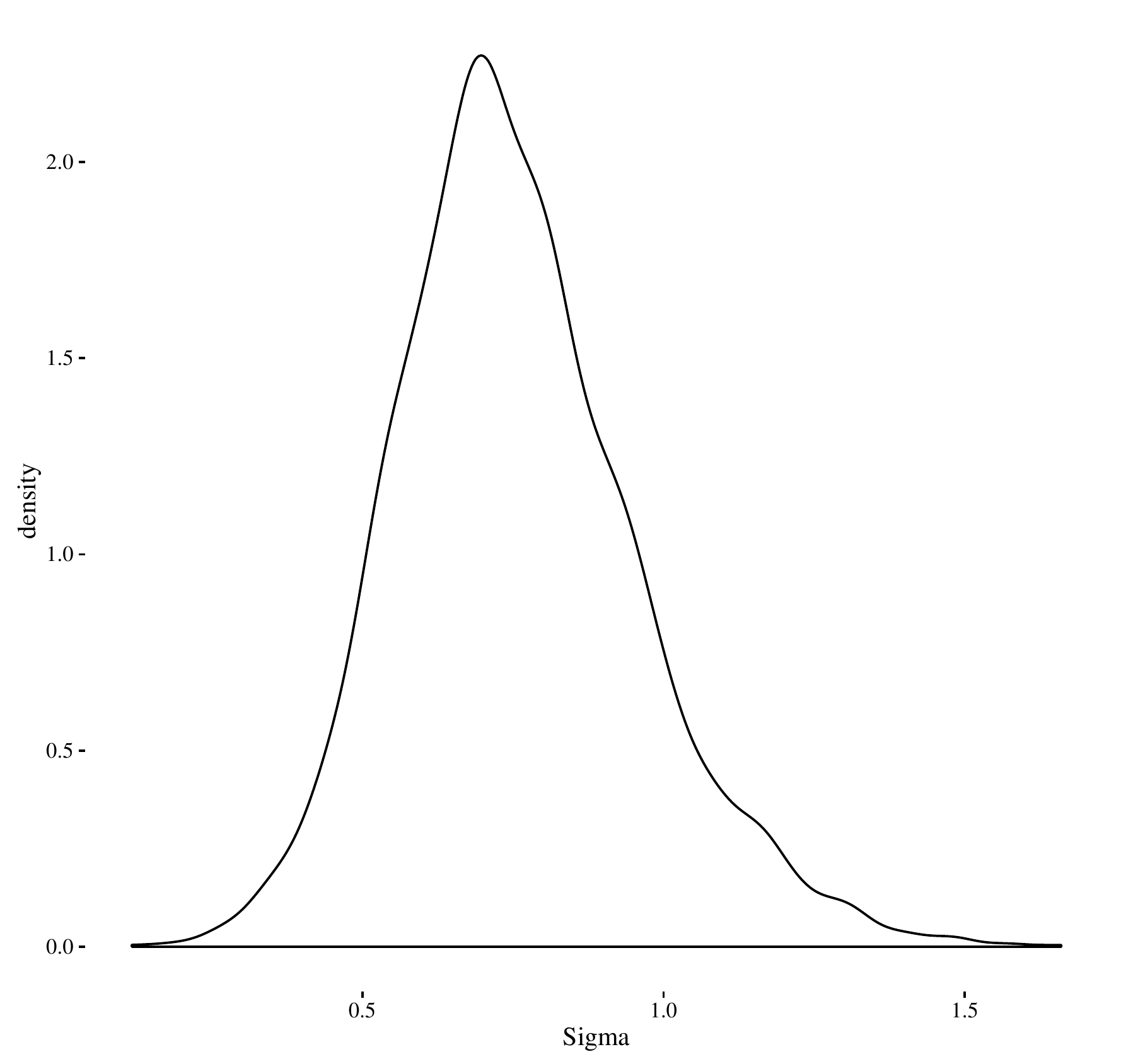}
       \caption{Density plot of the variance estimates for the random intercepts for subjects produced by the Bayesian model, behavioral task}
         \label{kailenvariance}
    \end{figure}
    
  \begin{figure}[H]
       \includegraphics[width=12cm]{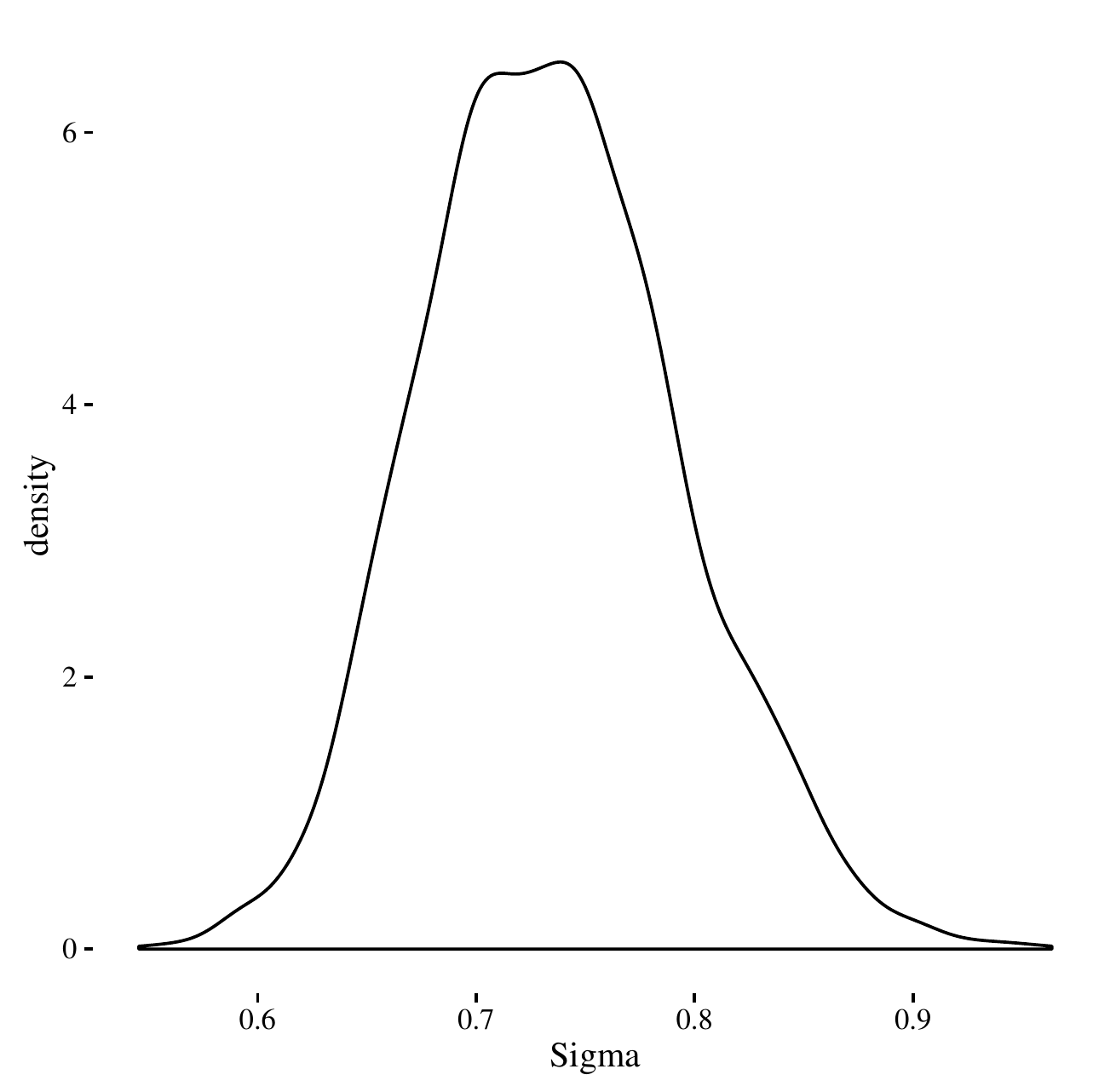}
       \caption{Density plot of the variance estimates for the random slope of primary stress by subjects produced by the Bayesian model, perception study}
         \label{ameliavariance}
    \end{figure}

\afterpage{%
\section{Figure Captions}
Figure 1. Density plot of the variance estimates for the random intercepts for subjects produced by the Bayesian model, behavioral task
\linebreak
\linebreak
Figure 2. Density plot of the variance estimates for the random slope of primary stress by subjects produced by the Bayesian model, perception study
\clearpage
}

\end{document}